\documentclass[lettersize,journal]{IEEEtran}
\usepackage{amsmath,amsfonts}
\usepackage{algorithmic}
\usepackage[ruled]{algorithm2e}
\usepackage{array}
\usepackage{textcomp}
\usepackage{stfloats}
\usepackage{url}
\usepackage{verbatim}
\usepackage{graphicx}
\usepackage{subfigure}
\def\BibTeX{{\rm B\kern-.05em{\sc i\kern-.025em b}\kern-.08em
    T\kern-.1667em\lower.7ex\hbox{E}\kern-.125emX}}
\usepackage{balance}
\usepackage{caption}
\begin{document}
\title{ Energy-efficient Caching and Task offloading for Timely Status Updates in UAV-assisted VANETs}
\author{Nan Hu,\mbox{\hspace{0.35cm}} Xiaoqi Qin, \mbox{\hspace{0.35cm}}Nan Ma, \mbox{\hspace{0.35cm}}Yiming Liu, \mbox{\hspace{0.35cm}} Yuanyuan Yao,\mbox{\hspace{0.35cm}} Ping Zhang,~\IEEEmembership{Fellow,~IEEE}

\thanks{N. Hu, X. Qin (\textit{corresponding author}), Y. Liu, N. Ma and P. Zhang are with Beijing University of Posts and Telecommunications, (e-mail: \{nanhu; xiaoqiqin; liuyiming; manan; pzhang\}@bupt.edu.cn).}
\thanks{Y. Yao is with School of Information and Communication Engineering, Beijing Information Science and Technology University, (e-mail: yyyao@bistu.edu.cn).}
}

\maketitle

\begin{abstract}
Intelligent edge network is maturing to enable smart and efficient transportation systems. 
In this letter, 
we consider unmanned aerial vehicle (UAV)-assisted vehicular networks where UAVs provide caching and computing services in complement with base station (BS). 
One major challenge is that vehicles need to obtain timely situational awareness via orchestration of ubiquitous caching and computing resources. 
Note that cached data for vehicles' perception tasks contains time-varying context information, 
thus freshness of cached data should be considered in conjunction with task execution to guarantee timeliness of obtained status updates. 
To this end, 
we propose a two-stage performance metric to quantify the impact of cache refreshing and computation offloading decisions on the age of status updates. 
We formulate an energy minimization problem by jointly considering cache refreshing, 
computation offloading and aging of status updates. To facilitate online decision making, 
we propose a deep deterministic policy gradient (DDPG)-based solution procedure and incorporate differentiated experience replay mechanism to accelerate convergence. 
Simulation results show that the performance of proposed solution is competitive in terms of energy consumption for obtaining fresh status updates.
\end{abstract}

\section{Introduction}

The advent of edge intelligence empowered B5G networks unlocks the potential for automotive industries, 
such as smart transportation system \cite{Xu:WCL:background}.
To reduce human intervention in critical operations such as path planning and obstacle avoidance, 
it's essential for autonomous vehicles to obtain real-time situational awareness of surroundings via intelligent orchestration of ubiquitous caching and computing capabilities \cite{Ning:TIT:vanet cache computing}.

This new paradigm involves information flows around a control-loop from the vehicle to edge server and back to the vehicle. 
For example, 
a vehicle continuously generates environment perception tasks and offload to edge server in short of on-board computing and caching resources.
Then edge server executes the task (e.g., simultaneous localization and mapping) based on its cached data (e.g., pre-built HD maps) and feedback status updates to the vehicle. 
Note that the cached data usually contains dynamic driving-related context information, 
which should be refreshed frequently. 
Therefore, 
the timeliness of obtained status updates is determined by task execution duration and freshness of cached computing data. 

In most existing work, 
the consecutively generated computation tasks are treated independently,
and the corresponding strategy design focuses on minimizing energy consumption \cite{Niu:green 6G} or task execution latency \cite{-J. Ku:TVT:UAV mobile cache computing}\cite{Peng:JSAC:UAV-assisted VANET}\cite{UAV-MEC}. 
In\cite{Peng:JSAC:UAV-assisted VANET}, 
Peng et al. studied multi-dimensional resource management for UAV-assisted VANETs. 
In \cite{UAV-MEC}, 
Hu et al. proposed a new architecture to assist  multi-modal-task offloading.
However, 
as for time-critical control scenarios considered in this letter, 
status updates obtained by executing computation tasks are temporally correlated.
Moreover, a stale status update is of less value in terms of the degree to which it represents reality. 
In this sense, 
we employ the concept of age of information (AoI) to quantify the timeliness of obtained status updates \cite{age}, 
where a larger value of age indicates that the status update is of less value for the accuracy of environment perception.  
The concept of AoI has been applied to investigate resource allocation strategies in edge computing systems \cite{Li:ICC:vanet age mec}
and cache updating systems \cite{Bastopcu:TWC:age cache}. 
Different from these existing work, 
we aim to investigate the closed-loop performance by characterizing intertwined relationship between cache refreshing and task execution, 
so as to strike a balance between the timeliness of obtained status updates and the corresponding required energy budget. 

In this letter, 
we consider an UAV-assisted vehicular network,
where UAVs are deployed to provide flexible caching and computing services in complement with base station. 
To obtain timely status updates in an energy-efficient manner, 
we develop an energy minimization problem by joint considering cache refreshing, task offloading, and aging of status updates at vehicles.
In the formulated problem, we propose a two-stage performance metric to characterize the temporal correlations among task generation, task execution and cache refreshing, 
which serves as a measure of age of obtained status updates. 
To achieve real-time decision making, 
we propose a differentiated experience replay based DDPG algorithm. 
Simulation results validate the effectiveness of our proposed solution.

\section{Modeling and problem formulation}
\subsection{Scenario Description}

\begin{figure}[h]
	\centering
	\includegraphics[width=0.50\textwidth]{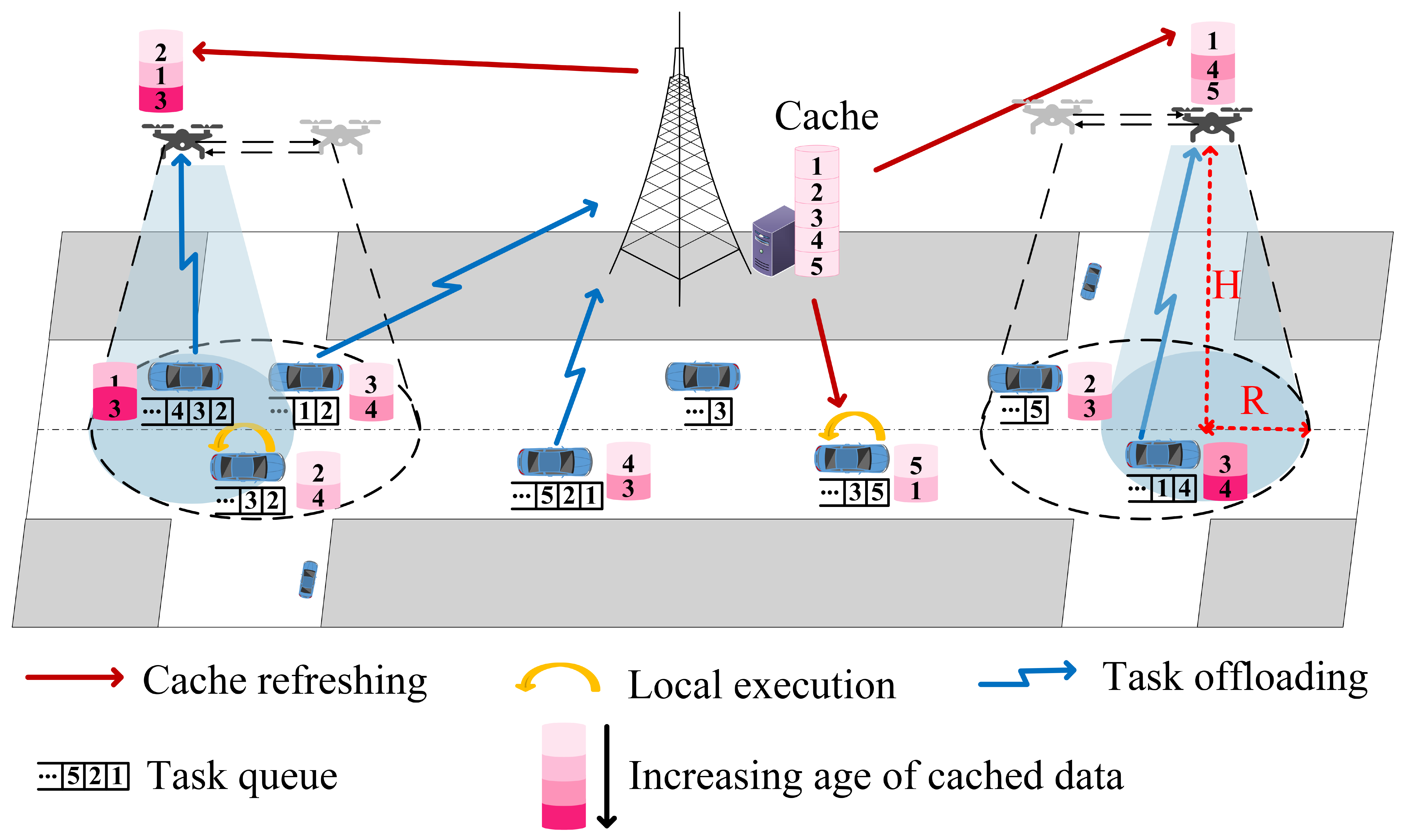}
	\caption{An UAV-assisted vehicular network.}
	\label{Fig:Network architecture}
\end{figure}

Consider an UAV-assisted vehicular network as shown in Fig. \ref{Fig:Network architecture}, 
which consists of a base station, 
a set of $\mathcal N$ vehicles and a set of $\mathcal M$ UAVs loitering over a specific segment of street with a constant speed and altitude. 
Denote $N=|\mathcal N|$ as the number of vehicles and $M=|\mathcal M|$ as the number of UAVs. 
Each UAV provides flexible cache-enabled edge computing service for vehicles within its coverage area,
which caches repetitively requested input data (e.g., high-resolution map) for timely task execution.
Denote  $\mathcal W$ as types of computation tasks.  
Upon each arrival of computation task $w$ at a vehicle, 
the task execution can be accomplished in three ways:
executed locally using data cached at a vehicle, 
offloaded to a nearby UAV using the ``flying" cache , 
or offloaded to base station which caches the most up-to-date data for all types of tasks.
Note that the size of cache at vehicles and UAVs is limited, 
thus the cached data has to be proactively selected and refreshed to collaborate with the task offloading decision. 
Table I lists notation used in this paper.
\begin{table}[htb]   
\begin{center}   
\caption{Notation}  
\label{table:1} 
\begin{tabular}{|c|c|} 

	\hline 
	Symbol                   &Definition\\
	\hline 
	$A^{veh}_{i,w}(t)$       & The age of cached data $w$ of vehicle $i$ at time slot $t$\\
	\hline  
	$A^{uav}_{j,w}(t)$       & The age of cached data $w$ of UAV $j$ at time slot $t$\\
	\hline  
	$\hat{A}_i(t)$           & The age of status updates of vehicle $i$ at time slot $t$\\
	\hline 
	$c^{veh}_{i,w}(t)$         & A binary variable to indicate whether or not  \\
	                           &data for task $w$ is cached at vehicle $i$ at time slot $t$\\
	\hline  
	$c^{uav}_{j,w}(t)$         & A binary variable to indicate whether or not \\
								&data for task $w$ is cached at UAV $j$ at time slot $t$\\
	\hline  
	$y^{veh}_{i,w}(t)$         & A binary variable to indicate whether or not data for task $w$\\
								& is updated or newly added to cache at vehicle $i$ at time slot $t$\\
	\hline  
	$y^{uav}_{j,w}(t)$         & A binary variable to indicate whether or not data for task $w$\\
								& is updated or newly added to cache at UAV $j$ at time slot $t$\\
	\hline  
	$x^{loc}_i(t)$                    & A binary variable to indicate whether or not \\
									&the task is executed locally at vehicle $i$ at time slot $t$\\
	\hline  
	$x^{mec}_{i,j}(t)$                    & A binary variable to indicate whether or not \\
										&vehicle $i$ offloads the task at time slot $t$\\
	\hline  
	$b_{i,j}(t)$                    & The bandwidth allocated to the vehicle $i$\\
	\hline 
\end{tabular}
\end{center}   
\end{table} 
%
\subsection{Timeliness of Status Updates}

To acquire timely situational awareness,
each vehicle continuously generates and executes tasks to obtain status updates of its surroundings.
Assume that sample-at-change strategy is adopted, 
in which the generation of task is event triggered and captures a change of status. 
Then the timeliness of status update at a vehicle is determined by i) the system time of the first unprocessed task left in its task queue (i.e.,  time duration from the task is generated until it is executed),  
and ii) the freshness of corresponding cached data for task execution. 
Therefore, 
it involves two set of decisions:
where tasks are executed (locally at vehicle, offloaded to UAV or base station)?
how to refresh the cached data at vehicles and UAVs?

\begin{figure}[h]
\centering
	\centering
	\includegraphics[width=0.48\textwidth]{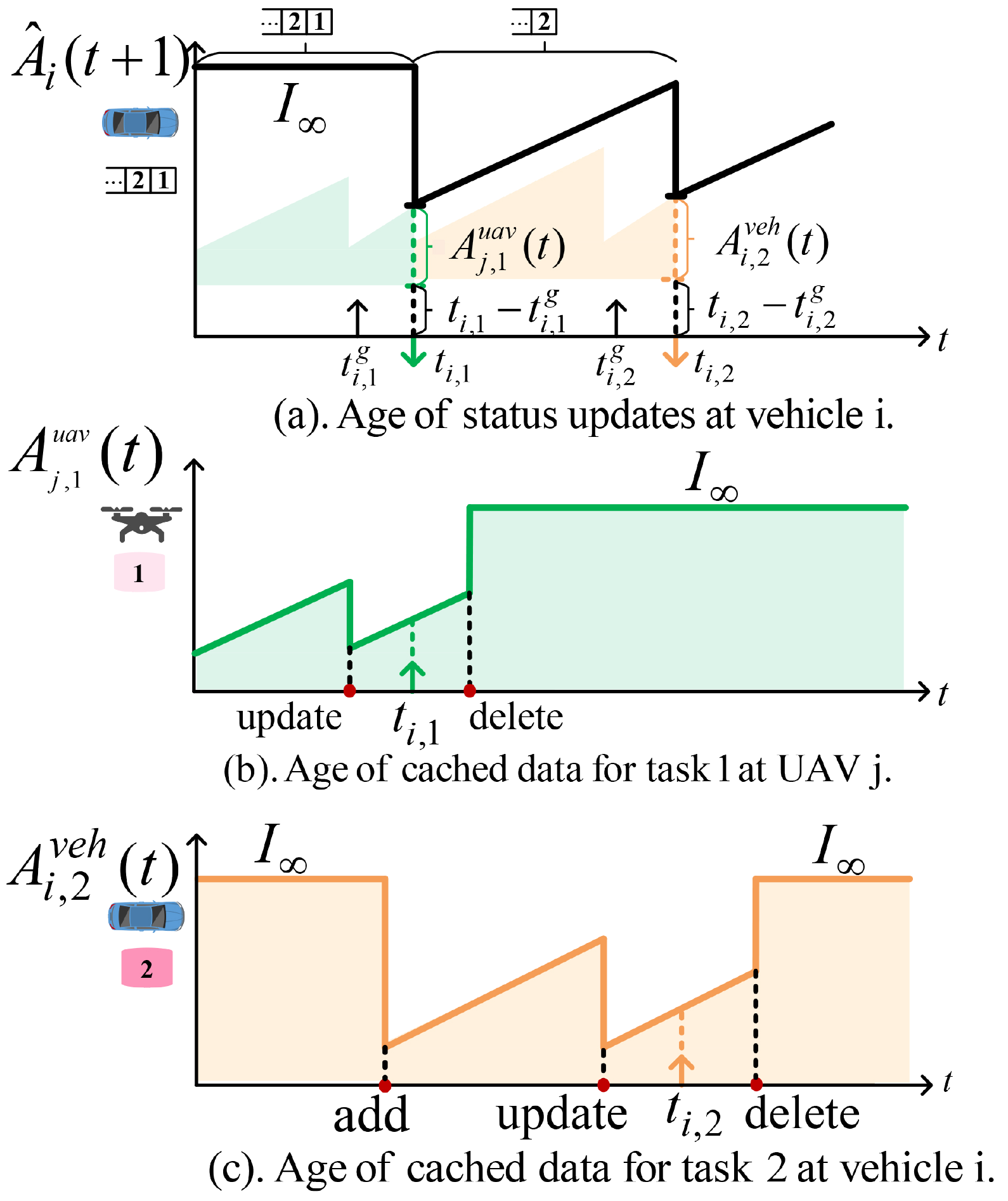}
	\caption{Evolution of age of status updates at vehicle $i$, age of cached data for task $1$ at UAV $j$ and age of cached data for task $2$ at vehicle $i$.}
	\label{Fig: Age of Status Updates }
\end{figure}

To optimize the timeliness of stats updates via aforementioned decisions, 
we propose two-stage performance metrics (i.e., age of cached data and age of status updates) based on the concept of age of information (AoI) \cite{age}. 
Note that since base station caches the most up-to-date data, the age of cached data is always 0. 
Denote $A^{veh}_{i,w}(t)$ as age of cached data for task $w$ at vehicle $i$ at time slot $t$, 
where $t\in\mathcal T=\{0,1,2,...,T\}$. 
Fig. \ref{Fig: Age of Status Updates } shows an example of evolution of $A^{veh}_{i,w}(t)$ . 
As shown in the figure, 
the cache refreshing has four possibilities: 
i)  $A^{veh}_{i,w}(t)$ starts at $1$ when it is newly added to cache;
ii) $A^{veh}_{i,w}(t)$ increases linearly with $t$ if it is not updated;
iii) $A^{veh}_{i,w}(t)$ drops to $1$ if it is updated;
iv) $A^{veh}_{i,w}(t)$ jumps to infinity if it is deleted due to limited storage space. 
Then we have: 
\begin{equation}
A^{veh}_{i,w}(t+1)=
\begin{cases}
1,                  &$newly add or update$,  \\
I_{\infty},         &$delete$, \\
A^{veh}_{i,w}(t)+1, &$otherwise$. \\
\end{cases}
\label{Eq: Age of Cache_vehicle}
\end{equation}
Denote $c^{veh}_{i, w}(t)$  as a binary variable to indicate whether or not data for task $w$ is cached at vehicle $i$ at time slot $t$. 
Denote $y^{veh}_{i,w}(t)$ as a binary variable to indicate whether or not data for task $w$ is updated or newly added to cache at vehicle $i$.
Then we have:
\begin{equation}
\begin{split}
y^{veh}_{i,w}(t) \leq c^{veh}_{i,w}(t+1),(i\in\mathcal N, w\in\mathcal W). 
\end{split}
\label{Eq: cache_vehivle}
\end{equation}
 
Then constraints (\ref{Eq: Age of Cache_vehicle}) can be transformed as follows:
\begin{equation}
	\begin{split}
		\hspace{-1mm}A^{veh}_{i,w}(t\hspace{-1mm}+\hspace{-1mm}1)\hspace{-1mm}=\hspace{-1mm}c^{veh}_{i,w}(t+1)[y^{veh}_{i,w}(t)\hspace{-1mm}+\hspace{-1mm}(1-y^{veh}_{i,w}(t))(A^{veh}_{i,w}(t)\\
		+1)]+(1-c^{veh}_{i,w}(t+1))I_{\infty},(i\in\mathcal N,w\in\mathcal W).
	\end{split}
	\label{Eq: Age of cached data_vehicle}
\end{equation}

Similarly, as for cache refreshing at UAV $j$, 
we have:
\begin{equation}
\begin{split}
y^{uav}_{j,w}(t) \leq c^{uav}_{j,w}(t+1),(j\in\mathcal M, w\in\mathcal W). 
\end{split}
\label{Eq: cache_UAV}
\end{equation}
\begin{equation}
\begin{split}
\hspace{-1mm}A^{uav}_{j,w}(t\hspace{-1mm}+\hspace{-1mm}1)\hspace{-1mm}=\hspace{-1mm}c^{uav}_{j,w}(t+1)[y^{uav}_{j,w}(t)\hspace{-1mm}+\hspace{-1mm}(1\hspace{-1mm}-\hspace{-1mm}y^{uav}_{j,w}(t))(A^{uav}_{j,w}(t)\\
+1)]+(1-c^{uav}_{j,w}(t+1))I_{\infty},(j\in\mathcal M,w\in\mathcal W).
\end{split}
\label{Eq: Age of cached data_UAV}
\end{equation}

Denote $\hat{A}_i(t)$ as age of status updates of vehicle $i$ at time slot $t$.
Fig. \ref{Fig: Age of Status Updates } shows an example of evolution of $\hat{A}_i(t)$ . 
As shown in the figure, 
the first task is generated at $t^g_{i,w}$ and $\hat{A}_i(t)$ increases linearly with $t$ before it is executed. 
Once task $1$ at vehicle $i$ is offloaded to UAV $j$ and processed at $t_{i,1}$, 
$\hat{A}_i(t)$ is reset to the sum of task $1$'s system time ($t-t^g_{1,w}$) and age of cached data for task $1$ at UAV $j$ ($A^{uav}_{j,1}(t)$).
Denote $t^g_{i,w}$ as the task generation time of task $w$ at vehicle $i$, 
then we have:
\begin{equation}
\hat{A}_i(t)=
\begin{cases}
t-t^g_{i,w}+A^{veh}_{i,w}(t), &$local execution$,  \\
t-t^g_{i,w}+A^{uav}_{j,w}(t), &$offloaded to UAV j$, \\
t-t^g_{i,w},    &$offloaded to BS$, \\
\hat{A}_i(t-1)+1, &$otherwise$. 
\end{cases}
\label{Eq: Age of Status Updates}
\end{equation}
To guarantee system-level timeliness of situational awareness,
we assume that age of status updates at each vehicle cannot exceed a threshold $A_{th}$. 
Then we have:
\begin{equation}
\begin{split}
\hat{A}_{i}(t) \leq A_{th},(i\in\mathcal N).
\end{split}
\label{Eq: Age of task's threshold}
\end{equation}
%
\subsection{Cache Refreshing Cost}

Denote  $C^{veh}_{i}$ as the storage space at vehicle $i$, 
$l_w$ as the size of input data for task $w$,
then we have:
\begin{equation}
\begin{split}
\mathop  \sum\nolimits_{w = 1}^{W} c^{veh}_{i, w}(t)l_w \le C^{veh}_i,(i\in\mathcal N, w\in\mathcal W). 
\end{split}
\label{Eq: vehicle's cache limit}
\end{equation}
Similarly, 
as for UAV $j$,  we have: 
\begin{equation}
\begin{split}
\mathop  \sum\nolimits_{w = 1}^{W} c^{uav}_{j, w}(t)l_w \le C^{uav}_j,(j\in\mathcal M, w\in\mathcal W). 
\end{split}
\label{Eq: UAV's cache limit}
\end{equation}

Based on constraints (\ref{Eq: Age of Status Updates}),
cached data at UAVs and vehicles should be refreshed as frequently as possible to improve the timeliness of status updates.
However, 
frequent cache refreshing brings extra energy consumption at these mobile terminals.
Denote $\theta$ (in $J$/bit) as the energy consumption for data fetching \cite{Nath:ITGCN:energy fetch cost}. 
Denote $\xi (t)$ as system-level energy consumption for cache refreshing at time slot $t$, 
then we have:
\begin{equation}
\begin{split}
\xi(t) = [\sum\limits_{i = 1}^{N} \sum\limits_{w = 1}^{W} y^{veh}_{i, w}(t)l_w +\sum\limits_{j = 1}^{M} \sum\limits_{w = 1}^{W} y^{uav}_{j, w}(t)l_w]\cdot \theta. 
\end{split}
\label{Eq: cache updating cost}
\end{equation}
%
\subsection{Computation Task Execution}

Denote $\mathcal{W}_i(t)$ as the types of computation tasks in vehicle $i$'s task buffer at time slot $t$. 
Note that only one task of the same type will be stored in the buffer. 
At each time slot $t$, 
vehicle $i$ executes the first unprocessed task in its buffer.
As task division is not considered here, 
a task can be executed in one of three ways:
locally at vehicle, 
offloaded to base station
or a nearby UAV. 

Denoted $x^{loc}_i(t)$ as a binary variable to indicate whether or not the task is executed locally at vehicle $i$.
Denote $x^{mec}_{i,j}(t)$ as a binary variable to indicate whether or not vehicle $i$ offloads the task.
Specifically, 
if $j=M+1$, 
it indicates whether vehicle $i$ offloads the task to BS;
if $j\leq M$, 
it indicates whether vehicle $i$ offloads the task to UAV $j$.  
Then we have:
\begin{equation}
	\begin{split}
		x^{loc}_i(t)+\sum\nolimits_{j=1}^{M+1}x^{mec}_{i,j}(t)\leq 1,(i\in\mathcal N).
	\end{split}
	\label{Eq: no task division}
\end{equation}

Note that task execution decisions and cache refreshing decisions are intertwined, 
since a task can be executed only when its input data has been cached. 
As for the first unprocessed task $w$ at vehicle $i$ ($w=\hspace{-1mm}\rm{max}\{k|t-t^g_{i,k},k\in\mathcal W_i(t)\}$), 
we have:
\begin{equation}
\begin{split}
x^{loc}_i(t) \leq c^{veh}_{i,w}(t),(i\in\mathcal N).
\end{split}
\label{Eq: execution limit}
\end{equation}
\begin{equation}
\begin{split}
x^{mec}_{i,j}(t) \leq c^{uav}_{j,w}(t),(i\in\mathcal N,j\in\mathcal M).
\end{split}
\label{Eq: offloading limit}
\end{equation}
%

\noindent\textbf{Local Execution:}
In the case when vehicle $i$ executes its first unprocessed task $w$ locally ($w\hspace{-1mm}=\hspace{-1mm}\rm{max}\{k|t-t^g_{i,k},k\in\mathcal W_i(t)\}$),  
denote $f_i^{loc}$ (in cycles/s) as computation capability of vehicle $i$. 
We assume that task execution must be completed within one time slot.
Denote $z_w$ as the required number of cycles for task $w$, 
while $\tau$ represents slot length,
then we have:  
\begin{equation}
\begin{split}
x^{loc}_{i}(t) \cdot \frac{z_w}{f_i^{loc}} \leq \tau, (i\in\mathcal N).
\end{split}
\label{Eq: local execution time}
\end{equation}

Denote $E^{loc}_i(t)$ as the corresponding energy consumption, and $\mu$ as the energy coefficient per CPU cycle \cite{Wen:INFOCOM:local execution cost}, 
we have:  
\begin{equation}
\begin{split}
E^{loc}_i(t)=\mu\cdot (f_i^{loc})^2\cdot z_w, (i\in\mathcal N).
\end{split}
\label{Eq: local energy consumption}
\end{equation}
\noindent\textbf{Task offloading:}
In the case when vehicle $i$ offloads its first unprocessed task to UAV or BS, 
denote $t^{tr}_{i,j}(t)$ as transmission duration and $t^{c}_{i,j}(t)$ as execution duration.
Assume that task processing must be completed within one time slot.
we have:
\begin{equation}
\begin{split}
x^{mec}_{i,j}\cdot [t^c_{i,j}(t)+t^{tr}_{i,j}(t)] \leq \tau,(i\in\mathcal N,j\in\mathcal M).
\end{split}
\label{Eq: total delay}
\end{equation}
As for execution duration $t^c_{i,j}(t)$, 
denote $f_{i,j}(t)$ (in in cycles/s) as the computing resources allocated to vehicle $i$ at UAV $j$ or BS ($j=M+1$), 
while $F^{max}_{j}$ represents the total CPU frequency.
Assume that the computing resources are equally divided among the offloaded tasks, 
then we have:
\begin{equation}
\begin{split}
f_{i,j}(t)\leq \frac{F^{max}_{j}}{\sum\nolimits_{i=1}^{N}x^{mec}_{i,j}(t)},(i\in\mathcal N,1\leq j\leq M+1).
\end{split}
\label{Eq: CPU constraints}
\end{equation}
\begin{equation}
\begin{split}
t^c_{i,j}(t)=\sum\nolimits_{j=1}^{M+1}x^{mec}_{i,j}(t)\frac{z_w}{f_{i,j}(t)},(i\in\mathcal N).
\end{split}
\label{Eq: MEC execution time}
\end{equation}

As for transmission duration $t^{tr}_{i,j}(t)$, 
denote $b_{i,j}(t)$ as the bandwidth allocated to vehicle $i$, 
while $B$ represents the total bandwidth available in the system. 
Then we have:
\begin{equation}
	\begin{split}
		\sum\nolimits_{i=1}^{N}\sum\nolimits_{j=1}^{M+1} b_{i,j}(t) \leq B.
	\end{split}
	\label{Eq: bandwidth limit}
\end{equation}

If vehicle $i$ offloads its task to BS (i.e., $x^{mec}_{i,M+1}(t)=1$),
the achievable task transmission rate can be obtained as:
\begin{equation}
	\begin{split}
		\hspace{-1mm}r^{BS}_{i,j}(t)=b_{i,j}(t)log_2(1\hspace{-1mm}+\hspace{-1mm}\frac{P^{tr} g_i(t)}{\sigma^2}), (i\in\mathcal N,j\hspace{-1mm}=\hspace{-1mm}M+1).
	\end{split}
	\label{Eq: BS transmission rate}
\end{equation}

where $P^{tr}$ is transmission power, 
$g_i(t)$ is the average channel gain between vehicle $i$ and BS, 
and $\sigma^2$ is noise power. 

Only if vehicle $i$ is within the coverage of UAV $j$, 
it can offload its task to UAV. 
At each time slot $t$,
given current coordinates of vehicle $i$ ($n^{x}_{i}(t),n^{y}_{i}(t)$), 
vehicle $i$'s speed $v^{veh}_i(t)$ (moving right is the positive direction),
coordinates of UAV $j$ ($u^{x}_{j}(t),u^{y}_{j}(t)$), 
UAV $j$'s speed $v^{uav}_j(t)$, 
flying height $H$ and the radius of UAV's projected coverage area $R$, 
the feasibility of offloading to UAV $j$ can be obtained by :
\begin{equation}
\begin{split}
\hspace{-2mm}d_{i, j}(t+1)\hspace{-1mm}=\hspace{-1mm}[\left( (n^{x}_{i}(t)\hspace{-1mm}+\hspace{-1mm}v^{veh}_i(t)\tau)\hspace{-1mm}-\hspace{-1mm}(u^{x}_{j}(t)\hspace{-1mm}+\hspace{-1mm}v^{uav}_j(t)\tau)\right)^2\\
+\left( n^{y}_{i}(t)-u^{y}_{j}(t)\right) ^{2}+H^2 ]^{\frac{1}{2}},(i\in\mathcal N,j\in\mathcal M).
\end{split}
\label{Eq: distance}
\end{equation}
\begin{equation}
	x^{mec}_{i,j}(t)d_{i,j}(t+1)\leq \sqrt{R^2+H^2},(i\in\mathcal N,j\in\mathcal M).
	\label{Eq: match constraints}
\end{equation}

We assume that the channel condition is constant in one time slot. 
Assume that vehicle $i$ has LoS view towards UAV $j$ with a given probability \cite{Al-Hourani:channel gain}:
\begin{equation}
\begin{split}
\hspace{-1mm}P^{los}_{i, j}(t)\hspace{-1mm}=\hspace{-1mm}\frac{1} {1+\gamma exp(-\psi [\zeta_{i, j}(t)-\gamma])},(i\in\mathcal N,j\in\mathcal M),
\end{split}
\label{Eq: task offloading delay}
\end{equation}
where $\gamma$ and $\psi$ are constant parameters determined by transmission environment. 
The elevation angle can be calculated as: 
\begin{equation}
\begin{split}
\hspace{-1mm}\zeta_{i, j}(t)=\frac{180}{\pi}sin^{-1}(\frac{H}{d_{i,j}(t)}).
\end{split}
\label{Eq: elevation angle}
\end{equation}
Then the channel gain between vehicle $i$ and UAV $j$ can be obtained as:
\begin{equation}
\begin{split}
\beta_{i,j}(t)=P^{los}_{i,j}(t)\frac{1}{\eta_1}\beta_0(d_{i,j}(t))^{-\alpha}+(1-P^{los}_{i,j}(t))\frac{1}{\eta_2}\\
\times\beta_0(d_{i,j}(t))^{-\alpha}, (i\in\mathcal N,j\in\mathcal M).
\end{split}
\label{Eq: channel gain in rate}
\end{equation}
where $\beta_0=(\frac{4\pi f_c}{c})^{-\alpha}$ is channel gain of unit distance, 
$f_c$ is the carrier frequency, 
$c$ is speed of light, 
$\alpha$ is path loss exponent, 
$\eta_1$ and $\eta_2$ ($\eta_2>\eta_1>1$) are the excessive path loss coefficients in LoS and NLoS cases, respectively. 
Then the achievable task transmission rate can be obtained as:
\begin{equation}
\begin{split}
\hspace{-2mm}r^{uav}_{i,j}(t)=b_{i,j}(t)log_2(1+\frac{P^{tr}\cdot \beta_{i,j}(t)}{\sigma^2}),(i\in\mathcal N,
j\in \mathcal M).
\end{split}
\label{Eq: vehicle-UAV  transmission rate}
\end{equation}

Denote $s_w$ as size of vehicle $i$'s first unprocessed task $w$, 
then the task transmission duration can be obtained as:
\begin{equation}
\begin{split}
t^{tr}_{i,j}(t)\hspace{-1mm}=\hspace{-1mm}\sum\nolimits_{j=1}^{M}\frac{x^{mec}_{i,j}(t)\cdot s_w}{r^{uav}_{i,j}(t)}\hspace{-1mm}+\hspace{-1mm}\frac{x^{mec}_{i,M+1}(t)\cdot s_w}{r^{BS}_{i,M+1}(t)},(i\in\mathcal N)
\end{split}
\label{Eq: transmission time}
\end{equation}

Then the energy consumption of task offloading can be obtained as:
\begin{equation}
\begin{split}
E^{off}_i(t)= P^{tr}\cdot t^{tr}_{i,j}(t),
(i\in\mathcal N).
\end{split}
\label{Eq: energy}
\end{equation}

\subsection{Problem Formulation}
We aim at minimizing the system-level energy consumption (including cache refreshing and computation task execution) over $T$ time slots. 
Then the problem can be formulated as:

\begin{center}
	\begin{tabular}{ l l }
		\bf OPT-P&               \\
		&$ \mathop {\min }\limits_{\scriptstyle \textbf{S}(t)} $    $\sum\nolimits_{t = 1}^T \left(\xi(t)+\sum\nolimits_{i=1}^{N}(E^{loc}_i(t)+E^{off}_i(t))\right)$ \\
		s.t
		& Timeliness of status updates: (\ref{Eq: Age of cached data_vehicle})(\ref{Eq: Age of cached data_UAV})(\ref{Eq: Age of Status Updates})(\ref{Eq: Age of task's threshold});\\
		& Cache refreshing costs: (\ref{Eq: cache_vehivle})(\ref{Eq: cache_UAV})(\ref{Eq: vehicle's cache limit})(\ref{Eq: UAV's cache limit})\\
		& Computation task execution: (\ref{Eq: no task division})-(\ref{Eq: transmission time})\\
	\end{tabular}
\end{center}

$\textbf{S}(t)$ is a set, 
which represents the set of all the variables that need to be decided in this letter.  
$\textbf{S}(t)=\{ y_{i,w}^{veh}(t),y_{j,w}^{uav}(t),c_{i,w}^{veh}(t), c_{j,w}^{uav}(t),x_i^{loc}(t),x_{i,j}^{mec}(t),{b_{i,j}}(t), 1\leq i\leq N, 1\leq j\leq M+1, 1\leq w\leq W\}$. 
In this formulation,
$y^{veh}_{i,w}(t)$, $y^{uav}_{j,w}(t)$, $c^{veh}_{i,w}(t)$, $c^{uav}_{j,w}(t)$, $x^{loc}_i(t)$ and $x^{mec}_{i,j}(t)$ are binary variables, 
$b_{i,j}(t)$ is a continuous variable. 
The formulated problem falls in the form of a mixed integer nonlinear program (MINLP), which is intractable.


\section{DRL-based  Caching and Task Offloading Algorithm}

In this section, 
we propose an online decision making approach for UAV-assisted vehicular networks, 
in which at each time slot $t$, 
the cache refreshing decisions ($y^{veh}_{i, w}(t)$, $y^{uav}_{j, w}(t)$, $c^{veh}_{i,w}(t)$, $c^{uav}_{j,w}(t)$), 
task execution decisions ($x^{loc}_i(t)$ and $x^{mec}_{i,j}(t)$), 
and bandwidth allocation decisions $b_{i, j}(t)$ are optimized in order to minimize system-level energy consumption. 
This can be achieved by transforming the formulated problem OPT-P into a MDP problem, 
which is defined as: 

\subsubsection{\textbf{State ($S^c$)}}

At time slot $t$, 
the system state is defined as the set of vehicles' and UAVs' coordinates,
and the age of status updates at vehicles, 
$s^c(t)=\{V_i(t), U_j(t), A^{veh}_{i,w}(t), A^{uav}_{j,w}(t), \hat{A}_i(t)\}$. 

\subsubsection{\textbf{Action ($A^c$)}}
At time slot $t$, 
BS needs to make decisions for cache refreshing, 
task execution and bandwidth allocation,   
$\hspace{80mm}a^c(t)\hspace{-1mm}=\hspace{-1mm}\{ y^{veh}_{i, w}(t)\hspace{-0.5mm},\hspace{-0.5mm} y^{uav}_{j, w}(t)\hspace{-0.5mm},\hspace{-0.5mm}c^{veh}_{i, w}(t)\hspace{-0.5mm},\hspace{-0.5mm} c^{uav}_{j,w}(t)\hspace{-0.5mm},\hspace{-0.5mm}x^{loc}_{i}(t)\hspace{-0.5mm},\hspace{-0.5mm}x^{mec}_{i,j}(t)\hspace{-0.5mm},\hspace{-0.5mm}b_{i, j}(t) \}$. 

\subsubsection{\textbf{Reward ($R^c$)}}

We employ the total energy consumption for cache refreshing and task execution (objective function of \rm{OPT-P}) as reward function $r^c(t)$, 
which is defined as: 
\begin{equation}
\begin{split}
\hspace{-2mm}r^c(t)=f\left(\sum\limits_{t = 1}^T (\xi(t)+\sum\limits_{i=1}^{N}(E^{loc}_i(t)+E^{mec}_i(t)))\right)-P.
\end{split}
\label{Eq:reward}
\end{equation}

where $f(\cdot)$ is the negative exponential function that acts as a normalization, 
which can be obtained by constraints (\ref{Eq: cache_vehivle})-(\ref{Eq: Age of Status Updates}), (\ref{Eq: cache updating cost})-(\ref{Eq: distance}) and (\ref{Eq: task offloading delay})-(\ref{Eq: energy}).
The penalty $P$ consists of constraints (\ref{Eq: Age of task's threshold})-(\ref{Eq: UAV's cache limit}) and (\ref{Eq: match constraints}), 
which prevents age threshold violation, 
cache overflow, and infeasible task offloading due to mobility of UAVs and vehicles.  

Define value of the $k$-th state $s^c_k$ at expected as expected long-term discounted reward under policy $\pi$ starting from $s^c_k$, 
we have: 
\begin{equation}
\begin{split}
\hspace{-2mm}V(s^c_k|\pi)=\mathbb{E}_\pi \left[ \sum\nolimits_{l=1}^{\infty} \gamma^{l-1}r^c(t+l)|s^c_k=s^c(t) \right].
\end{split}
\label{Eq:DDPG_V_value}
\end{equation}
Then the state-action-value function can be obtained as:
\begin{equation}
\begin{split}
\hspace{-2mm}Q(s^c_k,a^c_k|\pi)=\mathbb{E}_\pi \left[ r^c(t+1)+\gamma V(s^c_{k+1})|\pi \right].
\end{split}
\label{Eq:DDPG_Q_value}
\end{equation}
Consider continuous state $s^c$ and action $a^c$, 
we define the following performance objective under a certain policy $\pi$. 
\begin{equation}
\begin{split}
J(\pi)&=\mathbb E_{\pi}\left[ Q(s^c,a^c|\pi) \right]\\
&=\int_{S^c}d^{\pi}(s^c)\int_{A^c}\pi(a^c|s^c)Q(s^c\,a^c|\pi)da^cds^c.
\end{split}
\label{Eq:objective-J}
\end{equation}
\begin{figure}[htbp]
	\centering
	\includegraphics[width=0.38\textwidth]{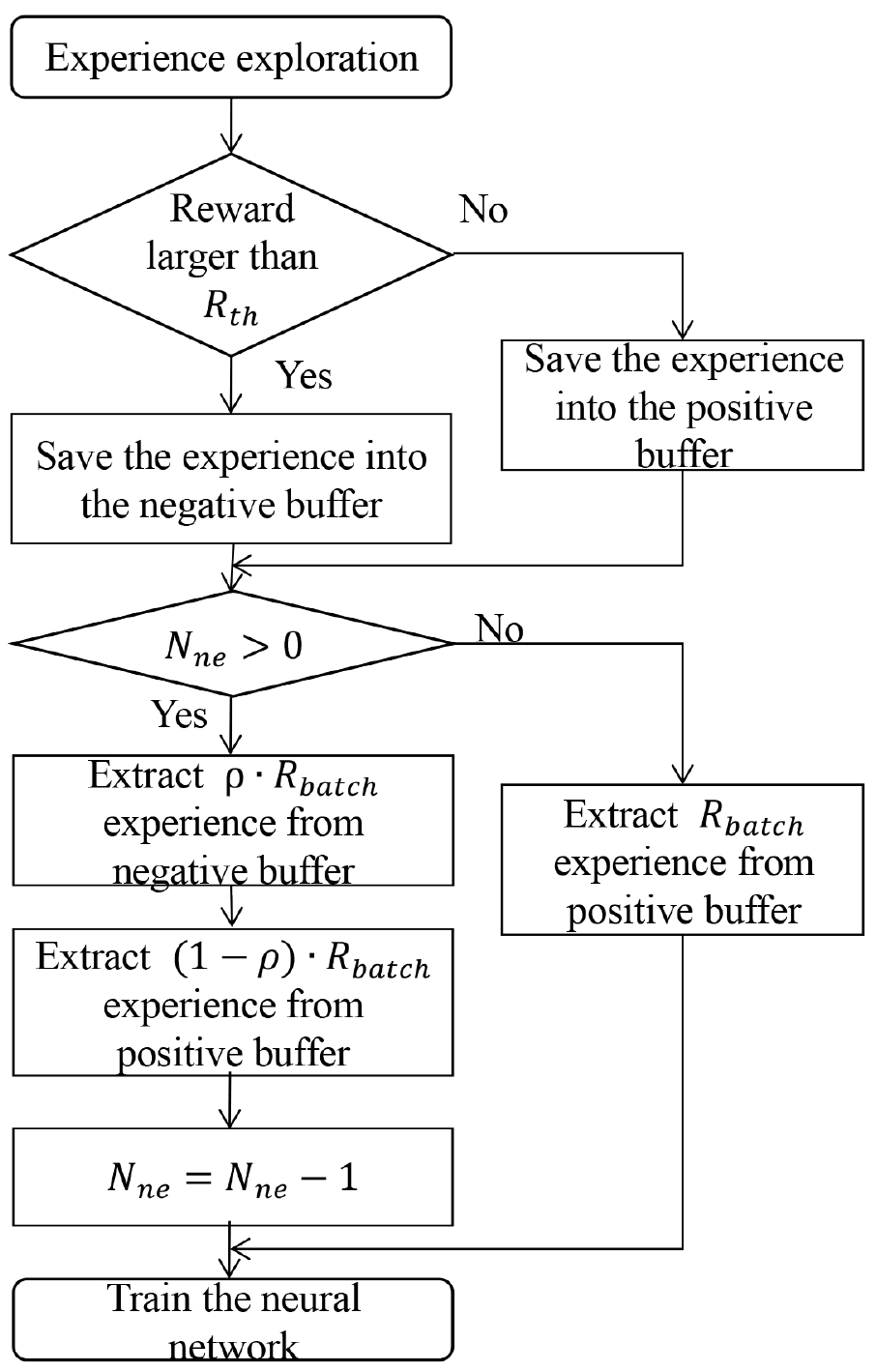}
	\caption{ The process of differentiated experience replay strategy.}\label{Fig:flow_chart}
\end{figure}

Due to randomness in task generation at each UAV,
the state transition probabilities are difficult to model. 
Therefore, 
we adopt a model-free reinforcement learning paradigm to learn and update the computation resource allocation policy.
Consider the continuous state space and action space, 
we propose a deep deterministic policy gradient (DDPG)-based algorithm to find solution for the aforementioned MDP problem.
DDPG follows actor-critic framework and combines ideas underlying the success of deep Q-network (DQN) and deterministic policy gradient to improve training efficiency. 
DDPG models contain consists of actor network I and critic network, while both of them contain two neural networks termed as online network and target network. 
The training process of DDPG is described as follows.

\textbf{Actor Network I Training Process}:
In training process of actor network I, 
policy gradient algorithm is employed to improve the parameterized policy by updating parameters in an iterative manner.
Define $\theta^{\mu}$ and ${\theta^{\mu'}}$ as parameters of online actor network and target actor network, respectively. 
Define $\mu(s^{\rm c}|\theta^\mu)$ as the deterministic policy parameterized by $\theta^\mu$.

According to deterministic policy gradient method, 
actor network outputs a deterministic action $a^{\rm c}(t)$ at epoch $v_k$ 
given state vector $s^{\rm c}(t)$, 
thus we have:
\begin{eqnarray}
a^{\rm c}(t) = \mu[s^{\rm c}(t)|\theta^\mu] + \mathcal{N}o(t) \; .
\label{Eq:action}
\end{eqnarray}
where $\mathcal{N}o(t)$ represents Uhlenbeck-Ornstein noise \cite{Yang}. 

Given the deterministic action, 
the performance objective function in Eq.~(\ref{Eq:objective-J}) can be rewritten as:
\begin{eqnarray}
J(\mu)& = \int_{S^{\rm c}}d^{\mu}(s^{\rm c})Q[s^{\rm c},\mu(s^{\rm c}|\theta^\mu)|\mu]ds^{\rm c}
\nonumber\\
& \hspace{-3em}
= \mathbb{E}_\mu\left[Q(s^{\rm c},\mu(s^{\rm c}|\theta^\mu)|\mu)\right] \; .
\label{Eq:objective-J2}
\end{eqnarray}

In \cite{Silver:DPG}, 
it has proved that the gradient of objective function under deterministic policy $\mu$ is equivalent to the expected gradient of Q function under policy $\mu$, then we have:
\begin{eqnarray}          
\nabla_{\theta^\mu}J(\mu)\hspace{-0.3em}=\hspace{-0.3em}\mathbb{E}_\mu\hspace{-0.3em}\left[\nabla_{\theta^\mu}\mu(s^{\rm c}|\theta^\mu)\nabla_{a^{\rm c}}Q(s^{\rm c},a^{\rm c}|\mu)|_{a^{\rm c}=\mu(s^{\rm c}|\theta^\mu)}\right].
\label{Eq:deterministic policy-objective}
\end{eqnarray}
To cope with correlation among data samples, 
DDPG employs experience replay mechanism where a memory buffer stores experience tuples $\{s^{\rm c}(v_{k}), a^{\rm c}(v_{k}), r^{\rm c}(v_{k+1}), s^{\rm c}(v_{k+1})\}$. 
Denote $\mathbb{B}$ as sampled mini-batch of data from the replay memory, 
and $Q(s^{\rm c},a^{\rm c})$ as value evaluation from critic network.
We employ Monte-Carlo method to estimate the expected gradient:
\begin{eqnarray}
\nabla_{\theta^\mu}J(\mu) \approx \frac{1}{|\mathbb{B}|}\sum_{i\in\mathbb{I}}\left[\nabla_{\theta^\mu}\mu(s^{\rm c}|\theta^\mu)|_{s^{\rm c}=s^{\rm c}_i}\times\right.
\nonumber\\
& \hspace{-11em}
\left.\nabla_{a^{\rm c}}Q(s^{\rm c},a^{\rm c}|\mu)|_{s^{\rm c}=s^{\rm c}_i,a^{\rm c}=\mu(s^{\rm c}_i|\theta^\mu)}\right].
\label{Eq:Monte-Carlo policy gradient}
\end{eqnarray}

\textbf{Critic Network Training Process}: 
The critic network evaluates actions generated by actor network I
using two neural networks termed as online Q network and target Q network. 
To approximate the value function, 
define ${\theta^Q}$ and ${\theta^{Q'}}$ as parameters of online Q network and target Q network, respectively. 
Define $Q(s^{\rm c}, a^{\rm c}|\theta^Q)$ as Q function estimated by online Q-network.

In the training process of critic network, 
$s^{\rm c}_i$ and $a^{\rm c}_i$ in mini-batch data $\mathbb{B}$ are fed into online Q network to output the evaluation Q value $Q(s^{\rm c}_i, a^{\rm c}_i|\theta^Q)$, where $i$ indicates the $i$-th sample in mini-batch of data. 
In target Q network, 
$s^{\rm c}_{i+1}$ and $\mu'(s^{\rm c}_{i+1}|\theta^{\mu'})$ from target actor network I are fed into neural network to 
generate target Q value $y_i$:
\begin{eqnarray}
y_i = r^{\rm c}_{i+1} + \gamma Q'[s^{\rm c}_{i+1}, \mu'(s^{\rm c}_{i+1}|\theta^{\mu'})|\theta^{Q'}]\; .
\label{Eq:target Q value}
\end{eqnarray} 
Then the mean square error $L$ for online Q network parameter update can be obtained as follows:
\begin{eqnarray}
L = \frac{1}{|\mathbb{B}|} \sum_{i\in\mathbb{I}}\left[y_i -Q(s^{\rm c}_i, a^{\rm c}_i|\theta^Q)\right]^2\; .
\label{Eq:mean square error}
\end{eqnarray} 

\textbf{Update Process}:
As for online actor network I,
the parameters are updated using gradient ascent method:
\begin{eqnarray}
\theta^\mu \leftarrow \theta^\mu + \rho^\mu\nabla_{\theta^\mu}J(\mu)\; .
\label{Eq:Act update}		
\end{eqnarray} 
As for online Q network in critic network, 
the parameters are updated using gradient descent method to improve estimation accuracy of Q function:
\begin{eqnarray}
\theta^Q \leftarrow \theta^Q - \rho^Q\nabla_{\theta^Q}L\; .
\label{Eq:Critic update}
\end{eqnarray}

As for target networks in actor network and critic network, 
the parameters are updated according to soft update method, 
where a small granularity is added to each step to stabilize the learning process:
\begin{eqnarray}
&\theta^{Q'} \leftarrow \chi \theta^Q + (1-\chi) \theta^{Q'},
\nonumber\\
&\theta^{\mu'} \leftarrow \chi \theta^\mu + (1-\chi) \theta^{\mu'}\; .
\label{Eq:soft update}
\end{eqnarray} 
where $0<\chi<1$ is the step size of soft update.

We propose a differentiated experience replay based DDPG algorithm to facilitate online decision making. 
Note that our proposed algorithm is problem-customized to make experience replay more efficient to achieve faster learning with better performance. 
More specifically, 
experience replay in traditional DDPG algorithms employs uniform sampling at random without considering the quality of experience. 
Considering the fact that an agent may learn more effectively from some transitions (including failures) than from others \cite{Schaul:Google:algorithm},
we classify the transitions into positive experience and negative experience based on a pre-fixed threshold of reward function values (denoted as $R_{th}$), 
and replay them in proportion to liberate agents from learning correlated transitions in the exact order they experienced. 

To distinguish between positive and negative experience,
we employ the lower bound value of reward function during convergence oscillation as the threshold ($R_{th}$).
Among total number of $ N_{step}$ steps within each episode, 
we choose the first $N_{ne} = N_{step} \cdot \chi$ steps to perform the aforementioned differentiated experience replay strategy, 
with $R_{ne}= R_{batch} \cdot \rho$ being the number of sampled negative experience. 
Fig. \ref{Fig:flow_chart} shows the process of differentiated experience replay strategy. 
Through numerous simulation, 
we found out that the most suitable range of coefficients are $\chi\in(0.05, 0.15)$ and $\rho\in(0, 0.2)$.

\begin{algorithm}[tb]
	\caption{DRL-based  Caching and Task Offloading Algorithm} 
	\begin{algorithmic}[1]
		\STATE {\bf Input:} State information at BS $s^{\rm c}(t)$, the fraction of sampled negative experience $\rho$, the number of step of sampling experience from negative experience buffer $N_{ne}$.\\
		\STATE {\bf Output:} Computation offloading decisions and cache refreshing decisions at BS $y^{veh}_{i, w}(t), y^{uav}_{j, w}(t),c^{veh}_{i, w}(t), c^{uav}_{j,w}(t),x^{loc}_{i}(t),x^{mec}_{i,j}(t),b_{i, j}(t).$
		\STATE {\bf Initialization:} 
		\STATE Randomly initialize online Q network $Q(s^{\rm c},a^{\rm c}|\theta^Q)$ and online actor network $\mu(s^{\rm c}|\theta^{\mu})$ with weight $\theta^Q$ and $\theta^{\mu}$.
		\STATE Initialize target network $Q'$ and $\mu'$ with weight $\theta^{Q'} \!\leftarrow \theta^Q$ and $\theta^{\mu'} \!\leftarrow \theta^{\mu}$.
		\STATE Initialize replay memory buffer.
		\FOR {$t=0,1,2...k...$}
		
		\STATE Get the task state information $s^{\rm c}(t)$.
		\STATE Perform action according to \eqref{Eq:action}.
		\STATE Calculate the reward according to \eqref{Eq:reward}.
		\IF {$r^{\rm c}(t+1)\leq R_{th}$}
		\STATE Store transition $\{s^{\rm c}(t),a^{\rm c}(t),r^{\rm c}(t+1),s^{\rm c}(t+1)\}$ to the positive experience memory buffer.
		\ENDIF
		\IF {$r^{\rm c}(t+1)>R_{th}$}
		\STATE Store transition $\{s^{\rm c}(t),a^{\rm c}(t),r^{\rm c}(t+1),s^{\rm c}(t+1)\}$ to the negative experience memory buffer.
		\ENDIF
		\IF {$N_{ne}>0$}
		\STATE Sample random $\rho*R_{batch}$ transitions from negative experience memory buffer, random $(1-\rho)*R_{batch}$ transitions from positive experience memory buffer, and $N_{ne}\leftarrow N_{ne}-1$.
		\ENDIF
		\STATE Calculate the Monte-Carlo policy gradient based on \eqref{Eq:Monte-Carlo policy gradient}.
		\STATE Calculate the mean square error according to  \eqref{Eq:target Q value}\eqref{Eq:mean square error}.
		\STATE Update online actor network according to \eqref{Eq:Act update}.
		\STATE Update online Q network according to \eqref{Eq:Critic update}.
		\STATE Update the target-networks according to \eqref{Eq:soft update}.
		\ENDFOR
	\end{algorithmic}
\end{algorithm}


\section{PERFORMANCE EVALUATION}

In this section, 
we present simulation results to demonstrate the performance of our proposed solution. 
We consider an UAV-assisted vehicular network where two UAVs with loiter height of $40$m provide flying caching and computation services for vehicles, 
while the communication range of an UAV is set as $100$m \cite{Peng:JSAC:UAV-assisted VANET}. 
Assume there are 5 types of tasks ($w=5$),
while task size and its required computation cycles follow uniform distribution with $s_w\in[100, 150]$ Kbits and $z_w\in[1\times 10^7, 1.5\times 10^7]$ cycles, respectively. 
The task generation frequency at each vehicle follows zipf distribution. 
The computation capability at UAV is $3\times 10^9$ cycles unit time.
The computation capability at vehicles follow uniform distribution with $f^{loc}_i\in[4.5\times 10^{8}, 5.5\times 10^{8}]$ cycles unit time. 
Each vehicle can cache data for one task, while UAV can cache data for three tasks.
The transmission power at vehicle is $1$ W. 
The energy coefficient of cache fetching is $10^{-8}$ J/bit \cite{Nath:ITGCN:energy fetch cost}, 
while energy coefficient for computing is $10^{-27}$ \cite{Wen:INFOCOM:local execution cost}. 
As for learning parameters,
the capacity of experience replay buffer is $10000$, which is equally divided into positive and negative buffers.
$\chi$ and $\rho$ are set as $0.1$. 

\begin{figure}[htbp]
	\centering
	\includegraphics[width=0.45\textwidth]{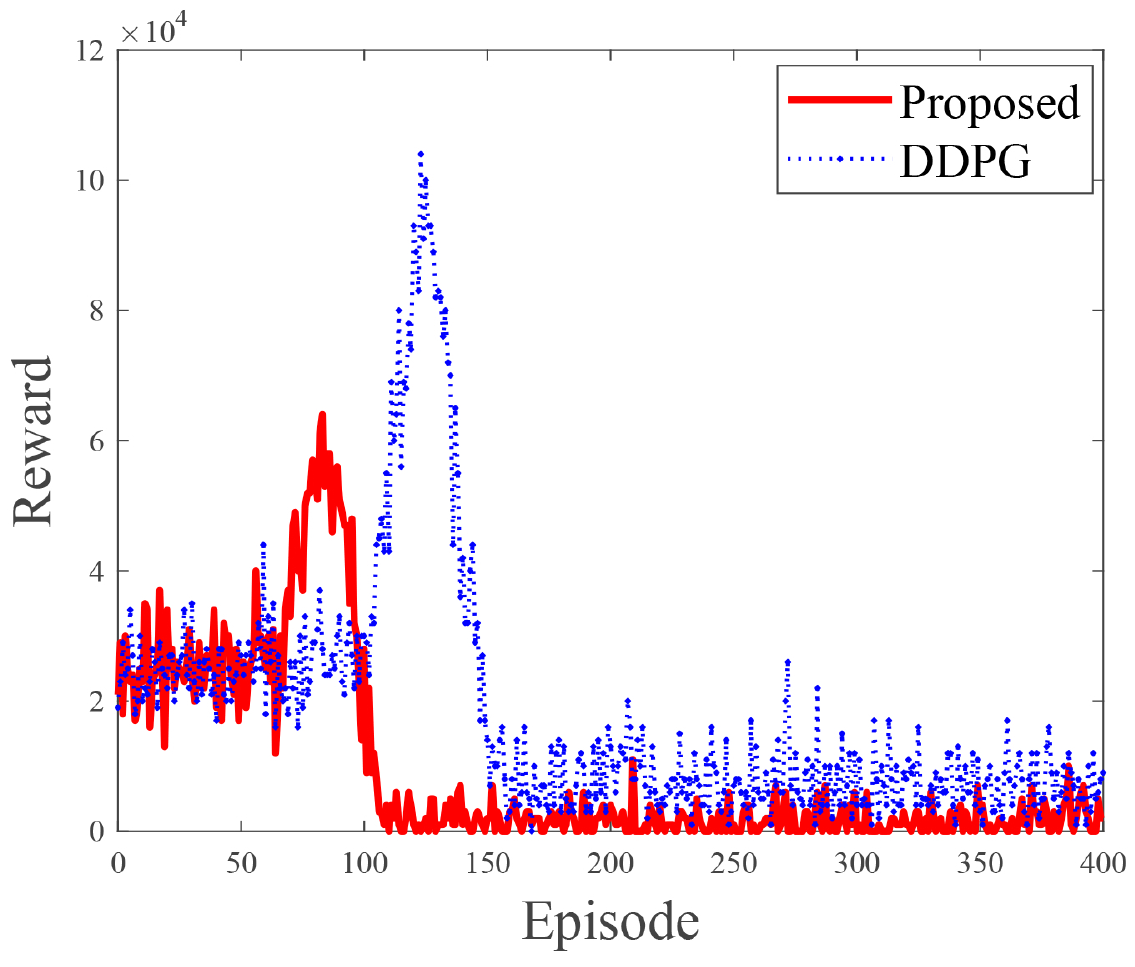}
	\caption{Reward values during training.}\label{Fig:reward}
\end{figure}

Fig. \ref{Fig:reward} compares the learning performance of our proposed solution with traditional DDPG algorithm.
As shown in the figure, 
our proposed solution achieves a better performance (in terms of minimizing energy consumption) with faster convergence rate.
It verifies the benefit of our proposed problem-customized differentiated experience replay.
\begin{figure}[htbp]
	\centering
	\includegraphics[width=0.45\textwidth]{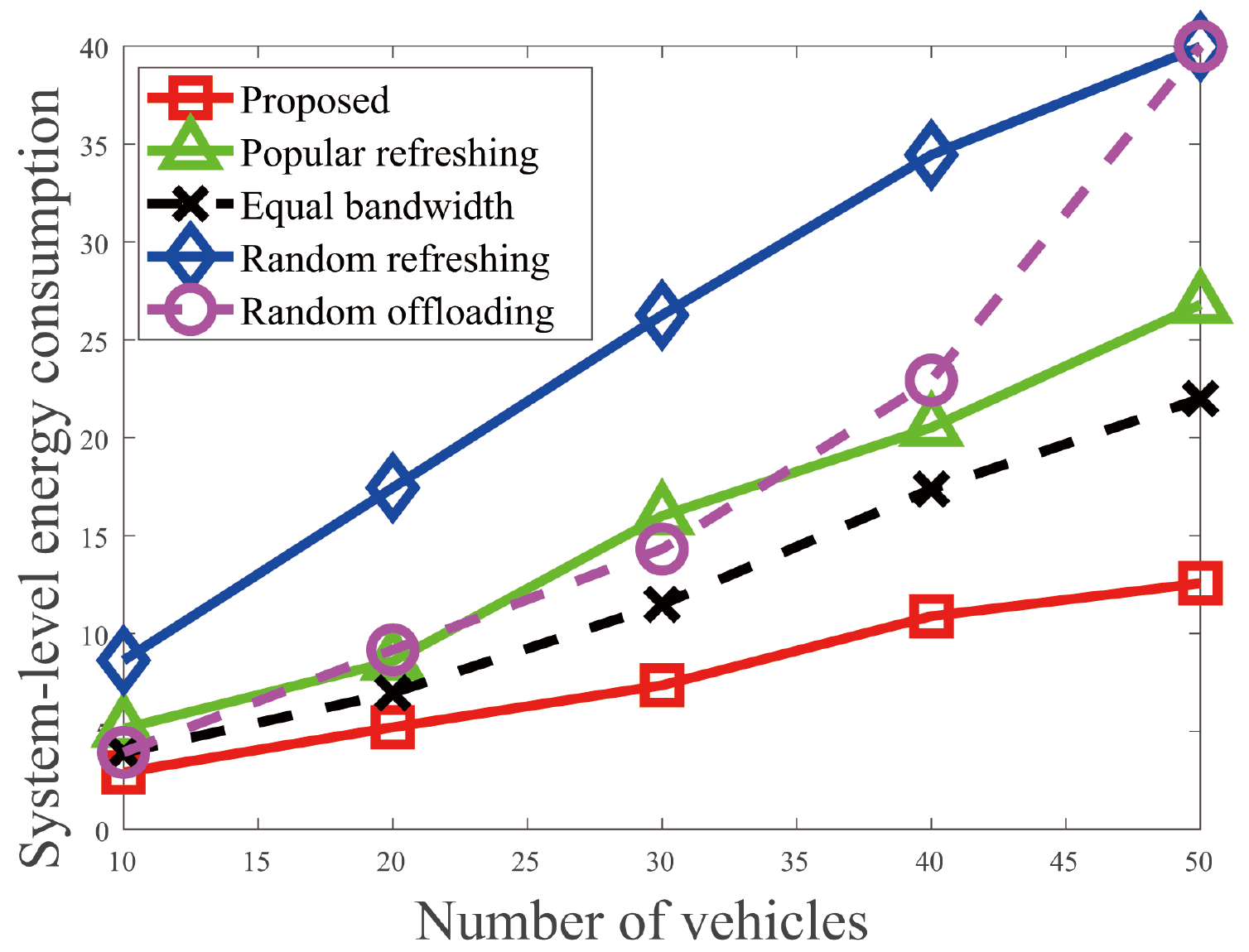}
	\caption{Energy consumption as increasing number of vehicles}\label{Fig:energy_vehicle}
\end{figure}

To demonstrate the performance benefit of our proposed solution, 
we employ four benchmarks.
As for ``Random refreshing'', 
the cached data at UAVs and vehicles is randomly refreshed with equal probability. 
As for ``Random offloading'', 
the offloading decision is made without considering caching availabilities at UAVs. 
In case when the required input data is not available, 
the task stays in the buffer and waits for next time slot. 
As for ``Popular refreshing'',
we employ task generation probability as task popularity, 
and the storage at UAVs and vehicles are refreshed to cache data for the most popular tasks.
As for ``Equal bandwidth'', 
the bandwidth is equally divided between offloaded tasks at each time slot. 

Fig. \ref{Fig:energy_vehicle} shows the trend of system-level energy consumption under five strategies as number of vehicles increases from $10$ to $50$. 
As shown in the figure, 
our proposed solution yields the best performance. 
Compared with ``Random refreshing'' and ``Random offloading'', 
our proposed solution greatly improves energy efficiency, which verifies the necessity of considering the interrelationship between caching and task execution. 
Moreover, 
the performance of our proposed solution is also better than ``Popular refreshing'' and ``Equal bandwidth'', 
which verifies the importance of a rigorous design of resource allocation optimization.
%
%

\section{CONCLUSION}

In this paper, 
we investigated an energy-efficient caching and task offloading strategy in UAV-assisted vehicular networks.
To quantify the timeliness of obtained updates, 
we employed the concept of age of information to bridge the gap between caching refreshing and task execution.
We formulated an energy consumption minimization problem by jointly considering cache refreshing, 
task execution and bandwidth allocation decisions. 
To realize fast decision making under stochastic task generations, 
we proposed a differentiated experience replay based DDPG algorithm. 
Simulation results demonstrated the performance benefit of our proposed solution in terms of energy efficiency and timeliness of status updates. 

\section*{Acknowledgments}
This paper is supported by National Key R\&D Program of China under Grant 2018YFB1800800, 
Beijing Natural Science Foundation under Grant No.L192033 and No.L192022.


%


\end{document}